\documentclass[12pt]{article}
\usepackage{amsmath2000}
\allowdisplaybreaks
\def\beq{\begin{equation}}
\def\eeq{\end{equation}}
\def\bea{\begin{eqnarray}}
\def\eea{\end{eqnarray}}

\def\be{\begin{equation}}
\def\ee{\end{equation}}
\def\ba{\begin{eqnarray}}
\def\ea{\end{eqnarray}}
\def\nn{\nonumber}

\def\Tr{{\rm Tr}}

\def\lam{\lambda}

\def\la{\langle}
\def\ra{\rangle}

\begin{document}

\pagestyle{plain} \setcounter{page}{1}
\begin{titlepage}

   \rightline{\small{\tt CALT-68-2411}}
   \rightline{\tt hep-th/0210153}

\vskip 1cm
\begin{center}
{\LARGE {SYM Description of SFT Hamiltonian \\[6pt]
   in a PP-Wave Background}}

\vskip 2cm {\large Jaume Gomis, Sanefumi Moriyama and Jongwon Park}

\vskip 1.2cm

\end{center}

\begin{center}
\emph{ California Institute of Technology 452-48, Pasadena, CA
91125}

\vskip .7cm {\tt gomis, moriyama, jongwon@theory.caltech.edu}

\end{center}
\vspace*{.5in}
\begin{center}
\textbf{Abstract}
\end{center}
\begin{quotation}
\noindent
We compute string field theory Hamiltonian matrix elements and compare
them with matrix elements of the dilatation operator in gauge
theory. We get precise agreement between the string field theory and
gauge theory computations once the correct cubic Hamiltonian matrix
elements in string field theory and a particular basis of states in
gauge theory are used. We proceed to compute the matrix elements of
the dilatation operator to order $g_2^2$ in this same basis. This
calculation makes 
a prediction for string field theory Hamiltonian matrix elements to
order $g_2^2$, 
which have not yet been computed. However, our gauge theory results
 precisely  match the results of the recent computation by Pearson et al.
 of the
order $g_2^2$ Hamiltonian matrix elements of the string bit model.

\end{quotation}
\vfil
\end{titlepage}

\newpage


\section{Introduction}

Recently, Berenstein, Maldacena and Nastase (BMN) \cite{BerensteinJQ}
have discovered a particular limit of the AdS/CFT correspondence --
known as the Penrose limit -- in which string theory can be solved to
all orders in $\alpha^\prime$ \cite{MetsaevBJ,MetsaevRE}.
The limiting geometry, which is the plane wave in \cite{BlauNE}, has
very different global properties compared to those of
$AdS_5\times S^5$.
Since the Penrose limit focuses on the geometry in the vicinity of a
particular massless particle in $AdS_5\times S^5$, which is very massive 
when measured by an $AdS_5$ observer,  information about the asymptotic
$AdS_5$ time-like boundary is lost in the limit, and the conformal
boundary of the plane wave degenerates to a null line
\cite{BerensteinSA}.
This peculiar degeneration makes it challenging to find a precise
holographic map between string theory and gauge theory.
On the other hand, the limit performed on the geometry can also be
realized on the dual gauge theory which leads BMN to conjecture that a
certain sector of ${\cal N}=4$ SYM captures the dynamics of string
theory in the plane wave background.

A crucial step in making the correspondence precise is to identify the
charges of string states in space-time with the charges carried by
gauge theory operators. The identification is given by
\be
\label{identify}
\frac{1}{\mu}H=(\Delta -J), \qquad \qquad \mu R^2P^{+}={J},
\ee
where $H$ is the generator of $x^+$ translations, $P^+$ is the
generator of $x^-$ translations, $\Delta$ is the generator of dilatations and
$J$ is a $U(1)_R$ charge. In the limit, the gauge theory truncates to
the BMN sector of operators carrying large $U(1)_R$ charge
$J\sim\sqrt{N}$  in the $N\rightarrow \infty$ limit.
In this double scaling limit we keep the following parameters fixed
\be
\label{para}
\lambda^\prime=\frac{g^2N}{J^2}=\frac{1}{(\mu p^+\alpha^\prime)^2},\qquad
g_2={J^2\over N}=4\pi g_s(\mu p^+\alpha^\prime)^2,
\ee
which effectively count quantum loops \cite{BerensteinJQ,GrossSU}
and non-planar corrections
\cite{KristjansenBB,BerensteinSA,ConstableHW}, respectively.

Even in the absence of string interactions, the BMN proposal makes a
remarkable prediction about the spectrum of anomalous dimensions of BMN
operators in the gauge theory in terms of the free string spectrum on
the plane wave.
This prediction was analyzed perturbatively in the regime where
$\lambda^\prime\ll 1$ in \cite{BerensteinJQ,GrossSU} and confirmed to
all orders using superconformal invariance in \cite{SantambrogioSB}.

In the light-cone gauge, in which the plane wave string theory can be
readily analyzed, the free Hamiltonian receives  string corrections
\be
\label{hamil}
H=H_2+g_2H_3+\cdots \ ,
\ee
which to lowest order result in  non-trivial transition amplitudes
between one and two-string states.
In \cite{ConstableHW} a concrete proposal was made which relates these
string theory transition amplitudes with three point functions of BMN
operators in the gauge theory.
Their proposal is therefore the first attempt to construct a
holographic map at the interacting level between string theory in a
plane wave and gauge theory.

This interesting proposal\footnote{Various aspects of this proposal
were considered in \cite{Kiem:2002xn,Huang:2002wf,Chu:2002pd,
Lee:2002rm,Chu:2002qj,Klebanov:2002mp,Huang:2002yt,Gursoy:2002yy,Chu:2002eu,
Lee:2002vz,Dobashi:2002ar,Janik:2002bd}.} was put to an explicit test
by Spradlin and Volovich in \cite{SpradlinRV} where some Hamiltonian
matrix elements were computed using the string field theory vertex
constructed  previously in \cite{SpradlinAR}.
Exact agreement with the proposal given in \cite{ConstableHW} was
reported.
Unfortunately both the field theory and string theory  computations
suffer from errors which when taken into account invalidate the
proposal.
On the field theory side, operator mixing is more important than
initially contemplated in \cite{KristjansenBB,ConstableHW}.
Three-point functions of single trace operators with order $\lambda'$
interactions are not conformally invariant, but the correct form is
restored after operator mixing is incorporated,
as shown in \cite{BeisertBB,ConstableVQ}.
On the string field theory side, the prefactor acting on the delta
functional overlap of three strings that enters the Hamiltonian $H_3$
\cite{SpradlinAR} has a  minus sign error, which was first
reported by Pankiewicz \cite{PankiewiczGS}.
Once matrix elements are recomputed with the corrected Hamiltonian
$H_3$, which we calculate in section $2$, agreement with field theory is lost.
Therefore, what is the correct holographic map between string theory
in the plane wave \cite{BlauNE} and ${\cal N}=4$ SYM at the
interacting level?

The most straightforward way to proceed, which was first advocated in
a paper by Gross, Mikhailov and Roiban \cite{GrossMH}, is to take
the identification (\ref{identify}) between the string field theory
Hamiltonian $H$ and the generator of scale transformations $\Delta -J$
in ${\cal N}=4$ SYM as the holographic map for all $g_2$.
This holographic map therefore identifies Hamiltonian matrix elements
in string field theory with those of the dilatation operator in gauge
theory.
In order to test this identification one must find an explicit map
between the Hilbert space of string states and the Hilbert space of
states in the gauge theory.
Moreover, in order for the comparison to be meaningful one must
compute the matrix elements of these operators in a basis in which the
Hilbert space inner product in gauge theory is the same as that in
string field theory.
The obvious inner product in the Hilbert space of string theory is the
familiar inner product where, for example, the one-string states are
orthogonal to two-string states.

In gauge theory,  the Hilbert space inner product is induced by the
matrix of two-point functions of BMN operators\footnote{In this
formula the various rows and columns describe single trace, double
trace, etc components. A more complete characterization of this
matrix is given in section $3$.}
\be
\label{metric}
|x|^{2\Delta_0}\la O_A \bar{O}_B \ra=G_{AB}+\Gamma_{AB}
\ln (x^2\Lambda^2)^{-1},
\ee
where $G_{AB}$ is the Hilbert space inner product and $\Gamma_{AB}$ is
the matrix of anomalous dimensions. Unlike with the usual Hilbert
space inner product in string field theory, which remains diagonal to
all orders in $g_2$, perturbative corrections in gauge theory induce
operator mixing at each order in $g_2$ in perturbation theory and the
Hilbert space inner product 
is no longer diagonal. Direct comparison with string field theory
calculations requires correcting for operator mixing
systematically, order by order in the $g_2$, expansion by making
$G_{AB}$ orthonormal via a change of basis.

In order to relate string theory to gauge theory calculations via
(\ref{identify}) we must calculate the matrix elements of the dilatation
operator.  It is straightforward to show that the matrix elements of
the dilatation operator between states created by BMN operators are
given by the matrix of anomalous dimensions\footnote{Since BMN
operators are BPS or  nearly BPS the contribution from the action of
$\Delta$ coming from the bare
dimension of the operator is cancelled by the contribution
from the R-charge in (\ref{twopt}).}
\be
\label{twopt}
\la O_A|(\Delta-J)|O_B \ra = \Gamma_{AB}.
\ee
Comparison of matrix elements of $H$ and $\Delta$ requires first
making  the gauge theory inner product  orthonormal order by order in 
perturbation theory.  
 We can accomplish this by finding a  new basis of operators
${\tilde O}_A=U_{AB}O_B$ such that they are orthonormal
\be
\label{ortho}
UGU^\dagger=1.
\ee
When $g_2=0$  the correct identification
between string states and gauge theory operators was given by
BMN. Namely, an $n$-string state is described by an $n$-trace operator. Once
$g_2$ corrections are taken into account this
identification has to be modified. 
Therefore, the precise mapping  between string field theory states
$|s_A\ra$ and gauge theory states $|\tilde{O}_A\ra$ when $g_2\neq 0$
is given by\footnote{In light-cone string field theory the canonical
normalization of states is the usual delta function normalization $\la
s_A^\prime|s_B^\prime\ra=p_A^+ \delta(p_A^+-p_B^+)=J_A
\delta_{J_A,J_B}$, so that
$|s_A^\prime\ra=\sqrt{J_A}|s_A\ra$. Therefore, when comparing string
field theory 
results with 
gauge theory results we will have to take into account this
normalization factor, since gauge theory states have unit  norm
\cite{BerensteinSA,ConstableHW}.}    
\be
\label{Hilbert}
|s_A \ra \rightarrow |\tilde{O}_A \ra =U_{AB}|O_B \ra, \qquad
\la s_A|s_B\ra = \la \tilde{O}_A|\tilde{O}_B \ra =\delta_{AB}.
\ee
In section $3$ we give an expression for the change of basis to order $g_2^2$.
Once the right basis is found, the holographic map reads
\be
\label{identi}
{1\over \mu}\la s_A|H|s_B \ra = \la \tilde{O}_A|(\Delta-J)|\tilde{O}_B
\ra =
(U\Gamma
U^\dagger)_{AB}.
\ee
The authors of \cite{GrossMH}, where this proposal was first made,
concluded that this proposal is equivalent to the one given in
\cite{ConstableHW} involving three point functions.
However, we  find that the two holographic maps are not the same and
that the one given in (\ref{identi}) is the correct one.

An important subtlety in the holographic map (\ref{identi}) is that
the orthonormalization procedure of the gauge theory inner product is not
unique, namely the transformation matrix $U$ that makes 
$G_{AB}$ orthonormal is not unique\footnote{This ambiguity was first
pointed out in \cite{GrossMH}.}.
We uniquely fix the form $U$ to order $g_2$ by demanding that the
dilatation operator matrix elements agree with the matrix elements of
the corrected string field theory Hamiltonian that we calculate in
section $2$. We then evaluate $U$ to order $g_2^2$ which via
(\ref{identify}) makes 
a non-trivial prediction for string field theory Hamiltonian matrix
elements to order $g_2^2$ which have not yet been evaluated.
Remarkably, this purely gauge theory result we present is reproduced
by the order 
$g_2^ 2$ string bit Hamiltonian {\cite{Verlinde:2002ig,Vaman:2002ka}
matrix elements recently computed in
\cite{PearsonZS}. 

The plan of the rest of the paper is as follows.
In section $2$ we revisit the string field theory Hamiltonian $H_3$
and point out that there is an incorrect relative minus sign in
\cite{SpradlinRV}\footnote{This sign has been corrected in a recent revision.}.
We recompute Hamiltonian matrix elements with the
corrected sign, which we will use in section $3$.
In section $3$ we fix the form of the change of basis in gauge theory
to order $g_2$ by demanding that (\ref{identi}) holds when the string
field theory matrix elements in section $2$ are used. We
also evaluate the dilatation operator matrix elements to 
order $g_2^2$ by making a particular choice of basis to order $g_2^2$
which allows us to make a calculable prediction by using (\ref{identi})
about the string field theory Hamiltonian matrix elements to that order. We
note that the final answer agrees with the recent
result in \cite{PearsonZS} found using the string bit formalism
\cite{Verlinde:2002ig,Vaman:2002ka}.
We conclude in section $4$.

\subsection*{Note Added}
While this work was being finished, the paper \cite{PearsonZS} appeared
on the archive. They also reexamine the duality proposal made in
\cite{ConstableHW} when the corrected string field theory Hamiltonian
is used and also conclude that once this is taken into account that the
proposal in  \cite{ConstableHW} no longer holds. Moreover, they also
adopt the philosophy in \cite{GrossMH} for comparing gauge theory and
string theory and also get agreement.

\section{SFT computation revisited}
Before going into our gauge theory computation, let us perform the
correct string field theory calculation that we will compare it to.
This also corrects some errors in the previous literature.
In string field theory, the Hilbert space is a direct product of
$\ell$-string states,
\be
{\cal H}=\oplus_\ell{\cal H}_\ell.
\ee
All these states are orthogonal with respect to each other and  they have
an orthonormal 
inner product.
The full Hamiltonian $H$, representing infinitesimal  evolution along $x^+$,
includes the freely propagating part, $H_2$, and an interaction part,
$H_3$ and so on,
\be
H=H_2+g_2H_3+\cdots.
\ee
In the plane wave background the freely propagating part $H_2$ is simply
the energy of an infinite collection of harmonic oscillators
$\alpha_n^i$
and the three-string
interaction part $H_3$ contains the delta functional overlap of three
strings 
with a
prefactor acting on it which is necessary to appropriately realize the
supersymmetry algebra.
The string states, which are dual to a certain class of two impurity
BMN operators when 
$g_2=0$, are given by
\begin{align}
|n\ra&=\alpha_n^{1\dagger}\alpha_{-n}^{2\dagger}|0,1\ra\\
|n,y\ra\ra&=\alpha_n^{1\dagger}\alpha_{-n}^{2\dagger}
|0,y\ra\otimes|0,1-y\ra\\
|y\ra\ra&=\alpha_0^{1\dagger}|0,y\ra\otimes\alpha_0^{2\dagger}|0,1-y\ra,
\end{align}
where the first state represents a single string and the other two 
 represent   two-string states.
Here $|0,y\ra$ is the one-string vacuum state carrying  a fraction 
$0<y<1$ of the
total longitudinal momentum $p^+$ of the multi-string state.
The Hamiltonian matrix elements of these states are given by
\begin{align}
\la n|H_3|m,y\ra\ra&
\sim\bigl(F_{(1)|m|}^+F_{(3)|n|}^++F_{(1)|m|}^-F_{(3)|n|}^-\bigr)
\bigl(\bar N^{(13)}_{|m|,|n|}-\bar N^{(13)}_{-|m|,-|n|}\bigr),
\quad\mbox{if  }mn>0\label{nHmy1}\\
\la n|H_3|m,y\ra\ra&
\sim\bigl(F_{(1)|m|}^+F_{(3)|n|}^+-F_{(1)|m|}^-F_{(3)|n|}^-\bigr)
\bigl(\bar N^{(13)}_{|m|,|n|}+\bar N^{(13)}_{-|m|,-|n|}\bigr),
\quad\mbox{if  }mn<0\label{nHmy2}\\
\la n|H_3|y\ra\ra&\sim F^+_{(3)|n|}
\bigl(F^+_{(1)0}\bar N^{(23)}_{0,|n|}
+F^+_{(2)0}\bar N^{(13)}_{0,|n|}\bigr),
\quad\mbox{for }\forall n\ne 0
\label{nHy}
\end{align}
where $F_{m(r)}$ comes from the prefactor and the Neumann matrices 
$\bar N^{(rs)}_{mn}$
come from the delta functional overlap.
The negative modes are related to the positive modes by\footnote{We
would like to thank A. Pankiewicz for informing us of the correct
factor of $i$ in the second formula.} (here $m,n>0$)
\begin{align}
\bar N^{(rs)}_{-m,-n}
&=-\bigl(U_{(r)}\bar N^{(rs)}U_{(s)}\bigr)_{m,n},\label{N=UNU}\\
F^-_{(r)m}&=i\bigl(U_{(r)}F^+_{(r)}\bigr)_m,\label{extrais}
\end{align}
with $\bigl(U_{(r)}\bigr)_{m,n}
=\delta_{m,n}\Bigl(\sqrt{m^2+(\mu p^+_{(r)}\alpha')^2}
-\mu p^+_{(r)}\alpha'\Bigr)/m$.
Note that in (\ref{extrais}) we have an extra factor of $i$ compared to
the original literature \cite{SpradlinRV}\footnote{More
explicitly, formulas (3.15) and (3.21) in 
the original version of \cite{SpradlinRV} should have an extra factor
of $i$. This 
invalidates the evaluation of (4.4) \cite{SpradlinRV}.}, which was first
 pointed out  
in \cite{PankiewiczGS}.
Using (\ref{N=UNU}) and (\ref{extrais}), we can show that both
(\ref{nHmy1}) and (\ref{nHmy2}) reduce to the same expression.

In order to compare these answers with
perturbative gauge theory we must analyze the $\mu\to\infty$ limit of
(\ref{nHmy1}), (\ref{nHmy2}) and (\ref{nHy}).
The $\mu\to\infty$ behavior of both the
prefactor and the Neumann matrices was evaluated in \cite{SpradlinRV}. 
In the $\mu\to\infty$ limit, 
(\ref{nHmy1}), (\ref{nHmy2}) and (\ref{nHy}) yield\footnote{The
precise overall numerical factor of the cubic string field theory
Hamiltonian is not known. It is fixed by comparing with the gauge
theory calculation in the next section.} 
\begin{align} \label{stringft1}
{1\over \mu} \la n|H_3|m,y\ra\ra&=  {\lam^\prime \over 2\pi^2 } (1-y) 
\sin^2(n\pi y),\\
\label{stringft2}
{1\over \mu} \la n|H_3|y\ra\ra&= - {\lam^\prime \over 2\pi^2} 
\sqrt{y(1-y)}\sin^2(n\pi y).
\end{align}
These corrected results invalidate the agreement previously found in the
literature\footnote{Also due to this extra $i$, the result proven 
in
\cite{Lee:2002vz} that the
prefactor reduces to energy difference between incoming and outgoing
states should be modified. Instead it reduces to the energy difference in
$\cos$ modes minus that in $\sin$ modes that appear in the worldsheet Fourier
decomposition.}.
In the next section we will use these results to test the holographic
proposal in (\ref{identi}) and will show that agreement is found for a
particular choice 
of basis.

\section{Orthonormalization and comparison with SFT}
In this section, we make a change of operator basis such that the
gauge theory inner product is orthonormal order by order in $g_2$.
We then compute the matrix elements of the operator $\Delta-J$ in this
basis.
An important subtlety in this procedure is that the basis change that
makes the Hilbert space inner product $G$ orthonormal is not unique.
The leading term at $g_2=0$ is uniquely fixed by the original correspondence
explained in \cite{BerensteinJQ} between single/double string states 
and single/double trace operators.
When $g_2\neq 0$ the field theory operators start mixing and the
matrix with which we make the inner product orthonormal is not unique,
due to the 
familiar ambiguity when diagonalizing a matrix.
We propose to fix this ambiguity in the change of basis by taking
seriously the proposal in (\ref{identi}) and demanding that exact
agreement with the corrected string field theory results computed in
the previous section is obtained.
Exact agreement is obtained for a unique choice of 
basis.
This particular choice 
 is the unique choice for which the transformation matrix
is symmetric and real to order $g_2$.
We then proceed to calculate to next to leading order, to order
$g_2^2$.
Here there is also an ambiguity in the orthonormalization procedure.
If we make the strong assumption that the transformation matrix is
still symmetric and real to this order we can uniquely determine the
transformation matrix to order\footnote{This choice is
compatible with the proposal in \cite{PearsonZS}.}  $g_2^2$.
Given these assumptions we can calculate explicitly the order $g_2^2$
matrix element between the orthonormal basis states which reduce when
$g_2=0$ to  single trace operators. This result gives a prediction
using the proposal (\ref{identi}) for light-cone string field theory
matrix elements to order $g_2^2$ involving single strings, which has
not been constructed.
Despite this incompleteness of light-cone string field theory, the matrix
elements we predict from the gauge theory computation exactly match
the order $g_2^2$ matrix elements computed recently in \cite{PearsonZS}
using the string 
bit formalism \cite{Verlinde:2002ig,Vaman:2002ka}.
It would be very desirable to understand more precisely the relation
between light-cone string field theory and the string bit formalism.

The particular set of gauge theory operators that we are interested
in are the following single trace and double trace operators\footnote{Here we 
are using the notation in \cite{BeisertBB}.}
\begin{align}
{\mathcal O}^J &= {1\over \sqrt{JN^J}}  \Tr Z^J, \\
{\mathcal O}^J_{(i)} &= {1\over \sqrt{N^{J+1}}} \Tr\left(\phi_i Z^J
\right), \\
{\mathcal O}^J_n &= {1\over \sqrt{JN^{J+2}}} \sum_{l=0}^J e^{2\pi
iln/J} \Tr\left(\phi_1 Z^l \phi_2 Z^{J-l}\right), \\
{\mathcal T}^{J,y}_p &=\; 
:{\mathcal O}^{y\cdot J}_p {\mathcal O}^{(1-y)\cdot J}:\;,\\ 
{\mathcal T}^{J,y} &=\; :{\mathcal O}^{y\cdot J}_{(1)}
{\mathcal O}^{(1-y)\cdot J}_{(2)}:\,.
\end{align}
We need to make the inner product $G$ appearing in  the matrix of
two-point functions in (\ref{metric}) orthonormal and eventually
compute the matrix 
elements of $\Gamma$ in the orthonormal basis (\ref{identi}).
Both matrices $G_{AB}$ and $\Gamma_{AB}$ have a systematic expansion
in powers of $g_2$
\begin{align}
G &= {\bf 1} + g_2 G^{(1)}+ g_2^2 G^{(2)} +{\mathcal O}(g_2^3),\\
\Gamma&=\Gamma^{(0)}+g_2\Gamma^{(1)}+g_2^2\Gamma^{(2)}
+{\mathcal O}(g_2^3).
\end{align}
In the following we split these matrices into $3\times 3$ blocks
representing matrix elements involving ${\mathcal O}^J_n$,
${\mathcal T}^{J,y}_p$ and ${\mathcal T}^{J,y}$ respectively.
The indices $(n,m,\cdots)$ denote the worldsheet momentum of the single
trace BMN operators like, for example, ${\mathcal O}^J_n$.
The double indices $(py,qz,\cdots)$ represent for example
 the worldsheet momentum
and light-cone momentum fraction of the double trace operators
${\mathcal T}^{J,y}_p$, while $(y,z,\cdots)$ represent the fraction of
momentum carried by the operator ${\mathcal T}^{J,y}$.
These matrices have been computed in
\cite{BeisertBB,ConstableVQ}.
They are given by\footnote{We would like to thank U. Gursoy for reminding us 
that the off-diagonal elements of the matrix of
two-point  functions involving 
double-trace operators are non-zero at order $g_2^2$.}:
\be \label{g}
G={\bf 1}+g_2\begin{pmatrix}
0&C_{n,qz}&C_{n,z}\\C_{py,m}&0&0\\C_{y,m}&0&0
\end{pmatrix}
+g_2^2\begin{pmatrix}
M^1_{n,m}&0&0\\0&\la ?\ra&\la ?\ra\\0&\la ? \ra&\la ?\ra
\end{pmatrix},
\ee
\be \label{gamma}
{\Gamma\over\lam^\prime} = \begin{pmatrix}
n^2\delta_{n,m}&0&0\\0&{p^2\over y^2}\delta_{p,q}\delta_{y,z}&0\\0&0&0
\end{pmatrix}
+g_2\begin{pmatrix}
0&\Gamma^{(1)}_{n,qz}&\Gamma^{(1)}_{n,z}\\
\Gamma^{(1)}_{py,m}&0&0\\
\Gamma^{(1)}_{y,m}&0&0
\end{pmatrix}
+g_2^2\begin{pmatrix}
nmM^1_{n,m}+{1\over 8\pi^2}{\mathcal D}^1_{n,m}&0&0\\0&\la ?\ra
&\la ?\ra \\0&\la ?\ra&\la ?\ra
\end{pmatrix}.
\ee
The explicit form of the matrix elements are summarized in the
Appendix and we denote by $\la ? \ra$ matrix elements that have not yet been
computed. Luckily, we will not need them for our computations. Finding
them is, however, an important enterprise since via the holographic
map (\ref{identi}) they predict yet unknown matrix elements in
string field theory, like the order $g_2^2$ Hamiltonian matrix element of a
two-string state\footnote{One should also compute, however, the mixing
between double and triple trace operators to get this result.}.

Let us now apply a linear transformation $U$ to make the
inner product orthonormal. We require that 
\beq
U G U^\dagger = {\bf 1}
\label{UgU}
\eeq
and solve this equation order by order.
We can expand $U$ in a power series in $g_2$ and express it as
\beq
U={\bf 1}+g_2U^{(1)}+g^2_2U^{(2)}+{\mathcal O}(g_2^3).
\eeq
As explained in the beginning of this section we restrict our
attention to symmetric and real matrices.
This is motivated in part by the fact that the symmetric and real choice
uniquely leads to exact agreement with string field theory via
(\ref{identi}) to order $g_2$ as we will see below.
Clearly, having a better understanding of why this choice works is
very desirable.
Therefore, by assuming that $U$ is a symmetric  and real matrix, we
need to solve 
(\ref{UgU}).
Solving this equation order by order we get 
\bea
&&U^{(1)}=-{1\over 2}G^{(1)}, \\
&&U^{(2)}=-{1\over 2}G^{(2)}+{3\over 8}\bigl(G^{(1)}\bigr)^2.
\eea
In the new orthonormal basis, $\Gamma$ is transformed to
\be
\tilde\Gamma=U \Gamma U^\dagger.
\ee
We can determine $\Gamma$ order by order in $g_2$ by expanding
\be
\tilde\Gamma=\tilde\Gamma^{(0)}+g_2\tilde\Gamma^{(1)}
+g_2^2\tilde\Gamma^{(2)}+{\mathcal O}(g_2^3).
\ee
The matrix of anomalous dimensions in the new basis is therefore
\bea
&&\tilde\Gamma^{(0)}=\Gamma^{(0)},\\
&&\tilde\Gamma^{(1)}=\Gamma^{(1)}-\frac12\{G^{(1)},\Gamma^{(0)}\},
\label{Gamma1}\\
&&\tilde\Gamma^{(2)}=\Gamma^{(2)}-\frac12\{G^{(2)},\Gamma^{(0)}\}
-\frac12\{G^{(1)},\Gamma^{(1)}\}
+\frac38\bigl\{\bigl(G^{(1)}\bigr)^2,\Gamma^{(0)}\bigr\}
+\frac14 G^{(1)}\Gamma^{(0)}G^{(1)}.
\label{Gamma2}
\eea
Using (\ref{g}) and (\ref{gamma}) we can evaluate (\ref{Gamma1}) to
be
\beq
\tilde\Gamma^{(1)}=\begin{pmatrix}
0&\tilde{\Gamma}^{(1)}_{n,qz}&\tilde{\Gamma}^{(1)}_{n,z}\\
\tilde{\Gamma}^{(1)}_{py,m}&0&0\\
\tilde{\Gamma}^{(1)}_{y,m}&0&0
\end{pmatrix},
\eeq
where
\bea
\label{gauge1}
&&\tilde\Gamma^{(1)}_{n,py}=\tilde{\Gamma}^{(1)}_{py,n}
=\lam^\prime{\sqrt{1-y}\over\sqrt{Jy}}{\sin^2(\pi ny)\over 2\pi^2},\\
\label{gauge2}
&&\tilde\Gamma^{(1)}_{n,y}=\tilde{\Gamma}^{(1)}_{y,n}
=-\lam^\prime{1\over \sqrt{J}}{\sin^2(\pi ny)\over 2\pi^2}.
\eea
We note that after using the proposed holographic map (\ref{identi})
that the gauge theory results (\ref{gauge1}), (\ref{gauge2}) match with
the string field theory results (\ref{stringft1}),
(\ref{stringft2})\footnote{As explained in a footnote in page 4, in
order compare the string field theory answer with the gauge theory
answer, one must divide the string result by $\sqrt{Jy(1-y)}$ so
that both string field theory states and gauge theory states have unit
 norm.}.

We can now make a prediction about the order $g_2^2$ matrix elements
in string field theory by using (\ref{identi}).
In order to do that we must calculate the matrix of
anomalous dimensions in the new basis to order $g_2^2$.
By using  (\ref{g}), (\ref{gamma}) we can perform the sums in
(\ref{Gamma2}) to get\footnote{The formulas we need to compute the required sums are
summarized in the Appendix.}
\bea
\tilde\Gamma^{(2)}
=\begin{pmatrix}
\tilde{\Gamma}^{(2)}_{n,m}&0&0\\0&\la ?\ra&\la ?\ra\\0&\la?\ra &\la ?\ra
\end{pmatrix},
\eea
where\footnote{Numerically, this expression is the same as the the one in
\cite{ConstableHW} for the non-nearest
neighbor genus 1 single-trace two-point function.}
\bea
\tilde{\Gamma}^{(2)}_{n,m}=\begin{cases}
\displaystyle
{\lam^\prime\over 32\pi^4}
\left({3\over nm}+{1\over (n-m)^2}\right)&\mbox{ if }n\ne m,-m\\[6pt]
\displaystyle
{\lam^\prime\over 16\pi^2}
\left({1\over 3}+{5\over 2\pi^2n^2}\right)&\mbox{ if }n=m\\[6pt]
\displaystyle
-{15\lam^\prime\over 128\pi^4n^2}&\mbox{ if }n=-m
\end{cases}
\eea
and $\la ?\ra$ are quantities that we cannot determine since the full matrix of
two-point functions has not been computed to order $g_2^2$.
We can nevertheless make the following prediction
\be
\label{predict}
{1\over \mu} \la n|H|m\ra\Bigr\vert_{g_2^2}=\tilde{\Gamma}^{(2)}_{n,m}.
\ee
We should note that this quantity has not been yet been computed in
light-cone string field theory. However, (\ref{predict}) exactly agrees 
with the 
recent proposal in \cite{PearsonZS} for the Hamiltonian matrix
elements of the string bit Hamiltonian.

\section{Discussion}

In this paper we have studied the gauge theory realization 
of string field
theory Hamiltonian matrix elements. The answer,
which was already anticipated in \cite{GrossMH}, is that these matrix
elements correspond to matrix elements of the dilatation
operator. Using the corrected string field theory results in section
$2$ we find a preferred basis of states which yields  agreement between gauge
theory and string theory calculations. Moreover, we make a prediction
using a gauge theory computation for the Hamiltonian matrix elements
of single string states to order $g_2^2$, which have not yet been
computed. We, however, report precise agreement with the recent string
bit \cite{Verlinde:2002ig,Vaman:2002ka} Hamiltonian calculation
presented in \cite{PearsonZS}. An outcome of the corrected string
field theory calculation in section $2$ is that the  the proposal of 
\cite{ConstableHW} no longer holds. This paper gives evidence that the
correct correspondence is between matrix elements of the string
Hamiltonian and the dilatation operator in the gauge theory. 
 This proposal suggests that the only observables that
can be holographically computed in the plane wave string theory are
gauge theory two-point functions
\cite{Verlinde:2002ig,GrossMH}. Nevertheless, operator mixing between  
multi-trace  operators contains information about higher point functions in 
 gauge theory. In string field theory, the Hamiltonian matrix elements 
compute the matrix of anomalous dimensions in the orthonormal basis via
(\ref{identi}), which can be
read from the gauge theory two-point functions.

The mapping between the string field theory and gauge theory Hilbert
spaces is non-unique. By comparing the calculation of the
dilatation operator matrix elements with
the corrected string field
theory Hamiltonian matrix elements in section $2$ we fixed the
ambiguity, which picks a particular basis of states in the gauge theory
to be identified with string states. However, it would be very
desirable to have a first principles 
explanation of why this choice is the correct one.
In the recent paper \cite{PearsonZS} the authors motivate this choice by
proving that the string bit Hamiltonian simplifies in this basis, that
is, it truncates at finite order in $g_2$. It would be desirable to
understand the uniqueness of the basis choice more directly. In this 
choice
of basis the gauge theory computation we present in section $3$
exactly agrees with the recent calculation \cite{PearsonZS}
performed at order $g_2^2$ using
the string bit \cite{Verlinde:2002ig,Vaman:2002ka} Hamiltonian.

A fascinating open problem is to understand more precisely the
relation between light-cone string field theory and the string bit
formalism. The advantage of the string bit formalism is, as shown
in \cite{PearsonZS},  that the Hamiltonian truncates at order
$g_2^2$. This however seems to raise a puzzle. The Hamiltonian matrix
elements truncate at order $g_2^2$, which via 
the map (\ref{identi}) predicts that matrix elements of the dilatation
operator in some basis truncate at order $g_2^2$ even though the
Hilbert space inner product $G_{AB}$ and the matrix of anomalous
dimensions $\Gamma_{AB}$  have corrections to all orders in $g_2$. It
would be very interesting to study this 
prediction in detail.
It still remains very desirable, however, to
explicitly construct the light-cone string field theory Hamiltonian at
order $g_2^2$ and explicitly verify that its matrix elements are
those computed in (\ref{predict}).

Light cone string field theory and the string bit model have some complementary
features. In the string bit formalism it is easier to compute the
$g_2^2$ corrections, while the same problem is notoriously difficult
in light-cone string field theory. On the other hand, in light-cone
string field theory we can systematically evaluate $1/\mu$
corrections,  which are hard to obtain in the string bit model.
Computing these corrections is crucial in extending the duality beyond
leading order in $\lambda^\prime$ in the gauge theory. In particular,
the results in (\ref{nHmy1}), (\ref{nHmy2}) and (\ref{nHy}) make
non-trivial predictions about ${\mathcal O}(\lambda^\prime)$
corrections to the matrix of anomalous dimensions via
(\ref{identi}). The factorization theorems 
 proven in \cite{Schwarz:2002bc,PankiewiczGS} will be very
useful in computing systematically all these corrections. It would
also be
very desirable to compute these corrections directly in gauge theory.

\subsection*{Acknowledgment}
We would like to thank Hirosi Ooguri and John Schwarz for useful
discussions and comments, and Peter Lee for collaboration in the early 
stage of this work. We would also like to thank Neil Constable,
Umut Gursoy, Daniel Freedman, Matthew Headrick  and Radu Roiban for 
comments on the 
original version of the paper.
J.~G.~is supported by a Sherman Fairchild Prize Fellowship. 
This research was supported in part by the DOE grant 
DE-FG03-92-ER40701.
S.~M.~was supported in part by JSPS Postdoctoral Fellowships for
Research Abroad H14-472.
 
\section*{Appendix}
In this appendix, we present the definition of several matrices
appearing in the main text and some useful formulae necessary for
computing the 
matrix elements of $\Gamma_{AB}$ in the new basis. They have been obtained 
from \cite{BeisertBB}.
\smallskip

{\bf\large Matrix elements}\qquad ($|m|\ne |n|, m\ne 0, n\ne 0,
p\in {\mathbf Z}, 0<y<1$)
\bea
\bullet\hspace{-0.5cm}
&&C_{n,py}=C_{py,n}=\frac{ y^{3/2}\sqrt{1-y}}{\sqrt{J}\pi^2}
\frac{\sin^2(\pi n y)}{(p-ny)^2} \\
&&C_{n,y}=C_{y,n}=-\frac{1}{\sqrt{J}\pi^2} \frac{\sin^2(\pi ny)}{n^2}\\
\bullet\hspace{-0.5cm}
&&M^1_{n,n}={1\over 60}-{1\over 24\pi^2n^2}+{7\over 16\pi^4n^4}\\
&&M^1_{n,-n}={1\over 48\pi^2n^2}+{35\over 128\pi^4n^4}\\
&&M^1_{n,m}={1\over 12\pi^2(n-m)^2}-{1\over 8\pi^4(n-m)^4}
+{1\over 4\pi^4n^2m^2}+{1\over 8\pi^4nm(n-m)^2}\\
\bullet\hspace{-0.5cm}
&&\Gamma^{(1)}_{n,py}=\Gamma^{(1)}_{py,n}
=\left({p^2\over y^2}-{pn\over y}+n^2\right)C_{n,py}\\
&&\Gamma^{(1)}_{n,y}=\Gamma^{(1)}_{y,n}=n^2C_{n,y}\\
\bullet\hspace{-0.5cm}
&&{\mathcal D}^1_{n,n}={\mathcal D}^1_{n,-n}
={2\over 3}+{5\over\pi^2n^2}\\
&&{\mathcal D}^1_{n,m}={2\over 3}+{2\over\pi^2n^2}+{2\over\pi^2m^2}
\eea
\smallskip

{\bf\large Useful summation formulae}\\
When one multiplies two matrices in (\ref{Gamma2}), the following
formulae are useful\footnote{Similar identities can also be found in
the Appendix of \cite{ConstableVQ}.}:
\bea
\nonumber&&\hspace{-1cm}\bullet\;\;\sum_{p,y}C_{n,py}C_{py,m}
=\frac{1}{J\pi^4}J\int_0^1dyy^3(1-y)\sin^2(\pi ny)\sin^2(\pi my)
\sum_{p=-\infty}^\infty\frac{1}{(p-nr)^2(p-mr)^2}\\
&&\hspace{-0.5cm}=\begin{cases}\displaystyle
{1\over 6\pi^2(n-m)^2}+{1\over 4\pi^4n^2m^2}
+{1\over\pi^4nm(n-m)^2}-{1\over 4\pi^4(n-m)^4}&\mbox{ if }n\ne m\\[6pt]
\displaystyle
{1\over30}-{1\over12\pi^2n^2}+{1\over2\pi^4n^4}&\mbox{ if } n=m
\end{cases}\\
\nonumber&&\hspace{-1cm}\bullet\;\;
\sum_y C_{n,y}C_{y,m}=\frac{1}{J\pi^4}J\int_0^1dy
{\sin^2(\pi ny)\over n^2}{\sin^2(\pi my)\over m^2}\\
&&\hspace{-0.5cm}=\begin{cases}
\displaystyle{1\over 4\pi^4n^2m^2}&\mbox{ if }n\ne m,-m\\[6pt]
\displaystyle{3\over 8\pi^4n^4}&\mbox{ if }n=m,-m
\end{cases}\\
\nn&&\hspace{-1cm}\bullet\;\;
\sum_{p,y}{p\over y}C_{n,py}C_{py,m}=\frac{1}{J\pi^4}J\int_0^1dy
y^2(1-y)\sin^2(\pi ny)\sin^2(\pi my)
\sum_{p=-\infty}^\infty\frac{p}{(p-nr)^2(p-mr)^2}\\
&&\hspace{-0.5cm}=\begin{cases}
\displaystyle
(n+m)\left\{{1\over 12\pi^2(n-m)^2}+{1\over 4\pi^4n^2m^2}
+{1\over 8\pi^4nm(n-m)^2}-{1\over 8\pi^4(n-m)^4}\right\}
&\mbox{ if }n\ne m\\[6pt]
\displaystyle{n\over 30}-{1\over 12\pi^2 n}+{7\over 8\pi^4n^3}&
\mbox{ if }n=m
\end{cases}\nn\\ \\
\nn&&\hspace{-1cm}\bullet\;\;
\sum_{p,y}{p^2\over y^2}C_{n,py}C_{py,m}
=\frac{1}{J\pi^4}J\int_0^1dyy(1-y)\sin^2(\pi ny)\sin^2(\pi my)
\sum_{p=-\infty}^\infty\frac{p^2}{(p-nr)^2(p-mr)^2}\\
&&\hspace{-0.5cm}=\begin{cases}
\displaystyle\frac{n^2+m^2}{12\pi^2 (n-m)^2}
+\frac{n^6+m^6-2nm(n^4+m^4)+n^3m^3}{4\pi^4n^2m^2(n-m)^4}
&\mbox{ if }n\ne m\\[6pt]
\displaystyle{n^2\over 30} +{3\over 2\pi^4n^2}&\mbox{ if }n=m
\end{cases}
\eea

\end{document}